\documentclass[aps,twocolumn,superscriptaddress,floatfix,prl]{revtex4-1}

\usepackage{lipsum}
\usepackage{graphicx}
\usepackage{amsmath,amssymb}
\usepackage{braket}
\usepackage{mathtools}
\usepackage{hyperref}
\usepackage{multirow}
\usepackage{color}
\usepackage[normalem]{ulem}
\usepackage{amsfonts}
\usepackage{float}
\usepackage{pdfpages} % include appendix
\makeatletter
\usepackage{subfigure}

\def\beq{\begin{equation}}
\def\eeq{\end{equation}}
\def\bea{\begin{eqnarray}}
\def\eea{\end{eqnarray}}

\AtBeginDocument{\let\LS@rot\@undefined}
\makeatother

\begin{document}

\title{Non-linear fluctuating hydrodynamics for KPZ scaling in isotropic spin chains}

%\title{Coupled stochastic non-linear equations for KPZ dynamics in spin chains  }

 \author{Jacopo De Nardis}
\affiliation{Laboratoire de Physique Th\'eorique et Mod\'elisation, CNRS UMR 8089,CY Cergy Paris
Universit\'e, 95302 Cergy-Pontoise Cedex, France}

\author{Sarang Gopalakrishnan}
%\affiliation{Department of Physics, The Pennsylvania State University, University Park, PA 16802, USA}
\affiliation{Department of Electrical and Computer Engineering, Princeton University, Princeton, NJ 08544, USA}

\author{Romain Vasseur}
\affiliation{Department of Physics, University of Massachusetts, Amherst, MA 01003, USA}

\begin{abstract}

 Finite temperature spin transport in integrable isotropic spin chains is known to be superdiffusive, with dynamical spin correlations that are conjectured to fall into the Kardar-Parisi-Zhang (KPZ) universality class. However,  integrable spin chains have time-reversal and parity symmetries that are absent from the KPZ/stochastic Burgers equation, which force higher-order spin fluctuations to deviate from standard KPZ predictions. We put forward a non-linear fluctuating hydrodynamic theory consisting of two coupled stochastic modes: the local spin magnetization and its effective velocity. Our theory fully explains the emergence of anomalous spin dynamics in isotropic chains: it predicts KPZ scaling for the spin structure factor but with a symmetric, quasi-Gaussian, distribution of spin fluctuations. We substantiate our results using matrix-product states calculations.

\end{abstract}
\vspace{1cm}

\maketitle
\paragraph{\textbf{Introduction} ---}
The Kardar-Parisi-Zhang (KPZ) equation arises as a coarse-grained description of many superficially disparate systems~\cite{KPZ, halpin2015kpz,Takeuchi2018}. Systems governed by the KPZ equation exhibit universal scale-invariant behavior, characterized by nontrivial critical exponents and exactly known scaling functions~\cite{Prahofer2004}. Typically, KPZ occurs in nonequilibrium dynamics subject to noise, also in quantum systems, see for example \cite{PhysRevX.7.031016,Fontaine2022,PhysRevLett.125.040603,2204.00070,Bernard2020,PhysRevLett.128.070401,PhysRevLett.124.236802}. The departure from equilibrium gives rise to an ``arrow of time'': indeed, in surface-growth problems governed by KPZ, growing and shrinking are inequivalent processes. Recently, a peculiar instance of KPZ scaling was discovered in the finite-temperature spin dynamics of the quantum Heisenberg spin chain~\cite{Ljubotina19, Ljubotina_nature, Ilievski18, GV19, NMKI19, Vir20,Keenan2022,ye2022universal,PhysRevE.100.042116}. This system is neither subject to noise nor out of equilibrium, so at first sight it is an unnatural candidate for KPZ scaling. Nevertheless, both the scaling exponents and the precise scaling function match KPZ expectations. The anomalous nature of spin transport at the Heisenberg point has been seen in experimental studies of solid-state magnets~\cite{Scheie2021} and ultracold gases~\cite{2107.00038}. At present, there is only a quantitative theory of the exponent~\cite{GV19, NMKI19, NGIV20, Bulchandani2021,PhysRevLett.123.186601} (which is argued to be universal for all integrable systems with $SU(N)$ or any other continuous nonabelian symmetries~\cite{Ilievski2021, MatrixModels, KP20, ye2022universal}), as well as a proposed mechanism for the scaling function~\cite{Vir20}. However, we lack a derivation and an understanding of the KPZ emergence from microscopic or hydrodynamic considerations. In addition, as we will discuss next, symmetry arguments preclude higher-order dynamical spin fluctuations in the Heisenberg spin chain from matching KPZ expectations~\cite{PhysRevLett.128.090604}. In this work, we argue for a modified version of the KPZ scenario that does respect the symmetries and can be derived from the underlying hydrodynamic of the model. We present numerical evidence that our scenario correctly captures dynamical spin fluctuations at finite temperatures. 

\emph{\textbf{Context}}.---It is helpful at this point to introduce the quantum spin-$1/2$ Heisenberg spin chain and describe its observed connections with KPZ. This model has the Hamiltonian 
 \begin{equation}\label{eq:XXX}
 H =  \sum_{x=1}^L \vec{S}_x \cdot \vec{S}_{x+1}, 
  \end{equation}
where $\vec{S} = \vec{\sigma}/2$ are spin-$1/2$ operators.
We are interested in the transport of spin fluctuations; for specificity we choose our quantization axis along $z$. Thus, $\sum_x S^z_x$ is conserved. We will typically work around the fully symmetric high-temperature state with $\sum_x S^z_x = 0$.

The key numerical observation~\cite{Ljubotina19} is that the finite-temperature dynamical spin structure factor $C^{zz}(x,t) \equiv \langle S^z_x(t) S_0^z(0) \rangle \to \chi t^{-1/3} f_{\mathrm{KPZ}}( \lambda_{\rm KPZ} x/t^{2/3})$, at large times $t$, where $\chi$ is the spin susceptibility, $\lambda_{\rm KPZ}$ is a non-universal constant, and $f_{\mathrm{KPZ}}(u)$ is the universal scaling function of the KPZ class \cite{Prahofer2004,Ferrari2006}. Since this behavior is expected at any non-zero temperature, it is convenient to study it at infinite temperature, and we will specialize to that case in what follows. Canonically, the scaling function $f_{\mathrm{KPZ}}$ arises in the study of the stochastic Burgers equation
 \begin{equation}\label{eq:burgers}
\partial_t \rho + \partial_x \left( \frac{\lambda}{2} \rho^2 -  D\partial_x \rho -  \xi \right) =0 ,
 \end{equation}
where $\rho(x,t)$ is a conserved density, $\lambda$ represents the strength of the non-linearity, $D$ is a diffusion constant, and $\xi$ is white noise with a strength related to $D$ by the fluctuation-dissipation theorem. It is a standard result that for a field $h$ that obeys the KPZ equation, the field $\partial_x h$ obeys the stochastic Burgers equation. In equilibrium, the Burgers-field correlation function $\langle \rho(x,t) \rho(0,0) \rangle$ (averaged over noise) is known to have the same form as the numerically observed $C^{zz}(x,t)$ for the spin chain.

This numerical observation might suggest a correspondence between the conserved densities $S^z_x(t)$ (or its coarse-grained version $m(x,t)$) in the spin chain and $\rho(x,t)$ in the Burgers equation, as $m(x,t)\sim\rho(x,t)$. However, this correspondence cannot be correct, because $\rho$ is inherently chiral, namely excess (deficit) density unbalances moves always to the right (left). This is also manifest from the distribution of fluctuations of $h$. Starting from an equilibrium initial condition, the probability distribution of $h(x,t) \equiv \int_{-\infty}^x \rho(x',t) dx'$ has the universal Baik-Rains form~\cite{baik2000limiting}, with a large finite skewness at all times. However, in the Heisenberg spin chain, this quantity is evidently symmetric between positive and negative current fluctuations, so its equilibrium distribution cannot be skewed: $P(J_m) = P(-J_m)$. 
In this letter we show how using more than one coupled Burgers modes solves such issues. Our results can be extended to any other integrable model with non-abelian symmetry, displaying super-diffusive charge transport.

  \begin{figure}[t!]
  \includegraphics[width=0.45 \textwidth]{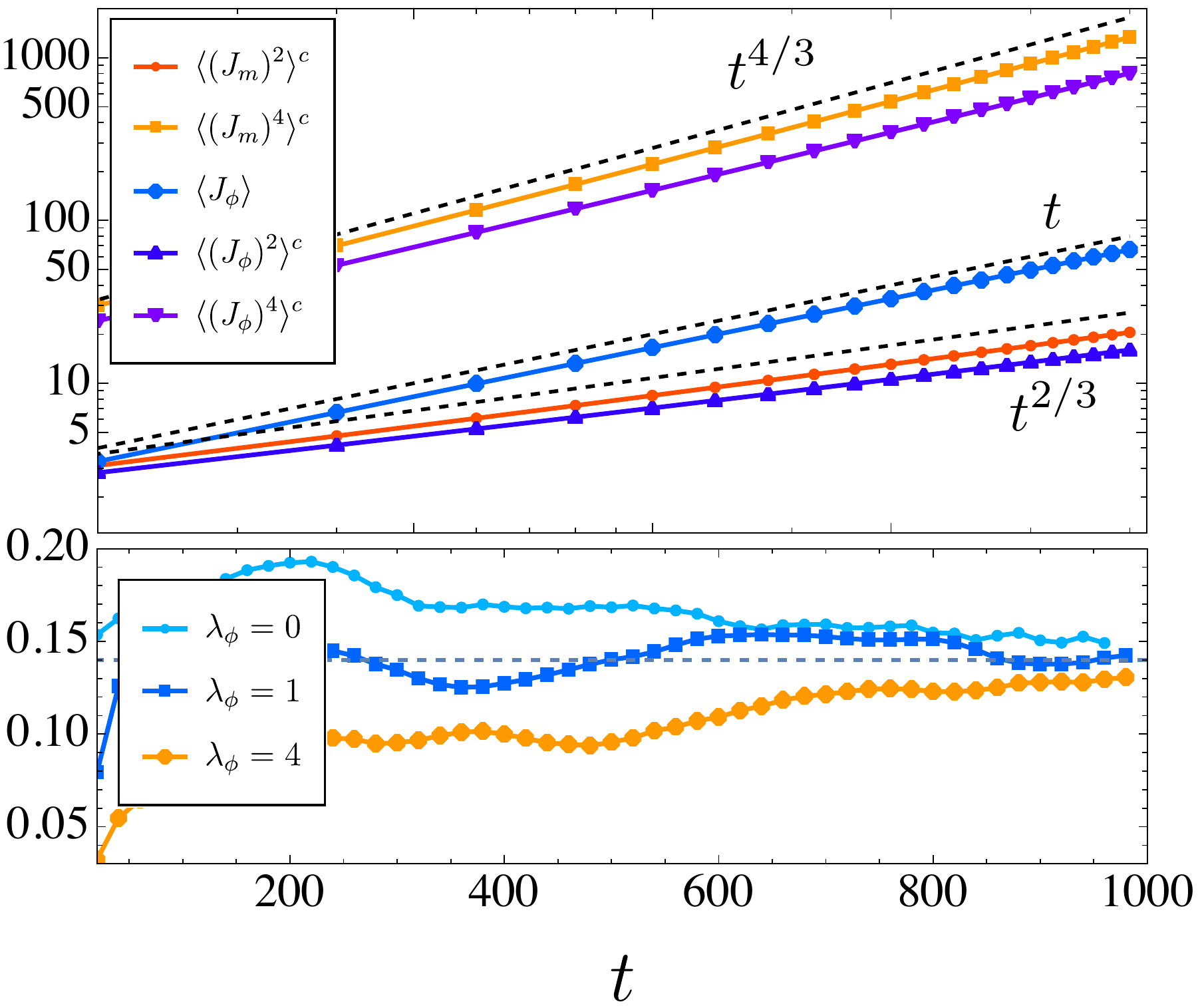}
  \caption{ {\bf Integrated current cumulants of the two-mode NLFH.} Log-Log plot of the cumulants of the integrated currents $J = \int_0^t \  j(x=0,t) dt$ for magnetization $m$ and for the chiral mode $\phi$ with $\lambda_\phi=0$, $\lambda_m=1$. The dashed lines are guides for the eyes indicating the different power laws. The odd cumulants of magnetization current are all 0 by symmetry. \textit{Bottom}: Kurtosis ${\rm Ku}_x =\langle x^4 \rangle/\langle x^2 \rangle^2 -3$ for the magnetization currents for different values of $\lambda_\phi$ in eq. \eqref{eq:coupledburger} and $\lambda_m=1$. The dashed line indicates half of Baik-Rains kurtosis $\sim 0.14$.   }\label{Fig:cumulants}
\end{figure}  
%A more quantitative way to see this obstruction is as follows. By the relation between $\rho$ and the KPZ height field $h(x,t) \equiv \int_{-\infty}^x \rho(x',t) dx'$, the probability distribution of the \emph{change} in the height field $P[h(0,t) - h(0,t=0)]$ at the origin is simply that of the integrated spin current $J_m=\int_0^t j_m(x=0,t') dt'$ across the origin in the time interval $(0,t)$~\cite{2107.00038, PhysRevLett.128.090604, PhysRevLett.128.160601, gopalakrishnan2022theory}. (More precisely, since we are dealing with a quantum system where the current operators at different times do not commute~\cite{levitov1993charge,doi:10.1063/1.531672,PhysRevB.51.4079}, $J_m$ is defined to be the total charge transfer through the origin up to time $t$.) 

\emph{\textbf{Hydrodynamics of giant quasiparticles}}.---We now introduce some key properties of the Heisenberg model that will feature in our logic. The model~\eqref{eq:XXX} is integrable and its hydrodynamics is given by the so-called Generalized Hydrodynamics (GHD)~\cite{PhysRevX.6.041065, PhysRevLett.117.207201, Doyon_notes}. This is a hydrodynamic theory for the evolution of the filling functions $n_s(\theta)/\sqrt{\chi_s(\theta)}$ (which represent the normal modes of the hydrodynamic theory and where $\chi_s$ are their susceptibilities) of the quasiparticles of the model. In general, $s$ and $\theta$ are some set of (respectively) discrete and continuous labels that denote each quasiparticle type; in an integrable system, the number of quasiparticles of each type is separately conserved. Physically, quasiparticles with $s > 1$ are bound states of elementary magnons \cite{Wang2018,Morvan2022}: { $s$ denotes the spatial extent and bare magnetization of the bound state, while the ``rapidity'' $\theta$ parametrizes momentum (energy) $k_s(\theta)$ ($\varepsilon_s(\theta)$), group velocity $v_s^{\rm eff}(\theta) = \partial \varepsilon_s(\theta)/\partial k_s(\theta)$, and bare spin $m_s = s$. The hydrodynamic normal modes move with velocity, $v_s^{\rm eff}(\theta)$ (which is an odd function of $\theta$ as expected) and therefore their occupation functions evolve with convective flow $\partial_t n_s(\theta) = - v_s^{\rm eff}(\theta)  \partial_x n_s(\theta)$. All quasiparticles are ballistic, however, the velocities $ v^{\rm eff}_s$ in a typical finite temperature state become tiny as their spin $s$ becomes large.  Indeed, quasiparticles with large spin correspond to large bound states, so-called ``giant'' quasiparticles, and can alternatively be seen as quasi-classical wave packets made up of Goldstone modes~\cite{NGIV20}. Since the Bethe vacuum is always ferromagnetic, the Goldstone mode dispersion is { quadratic $\varepsilon \sim k^2$}, so bound states of size $s$ have a characteristic { velocity $v^{\rm eff}_s \sim k_s \sim s^{-1} $}. When quasiparticles are present at finite density, the magnetization of the lower-$s$ quasiparticles is screened by the higher-$s$ ones, so spin transport is determined by giant quasiparticles with $s \to \infty$. Therefore, above a state with net magnetization density $m \ll 1$, fluctuations of the magnetization are primarily due to quasiparticles of size $s \sim 1/m$. 
 It is useful to introduce an external field $h = |m|/\chi$, with $\chi$ the spin susceptibility: giant quasiparticles are the ones with spin $s = \xi/h$ and rapidity $\theta = u/h$ with $h \ll 1$ and $\xi$ and $u$ real variables (such a semi-classical limit of quasiparticles was also employed recently in \cite{NGIV20,2303.16932}). In this limit, the sum over all the quasiparticle types becomes $\sum_s \int d\theta g_s(\theta ) \to h^{-2} \int_0^\infty d\xi \int_{-\infty}^{+\infty} du \, g_{s/h}(u/h) $.   
\begin{figure}[t!]
  \includegraphics[width=0.45\textwidth]{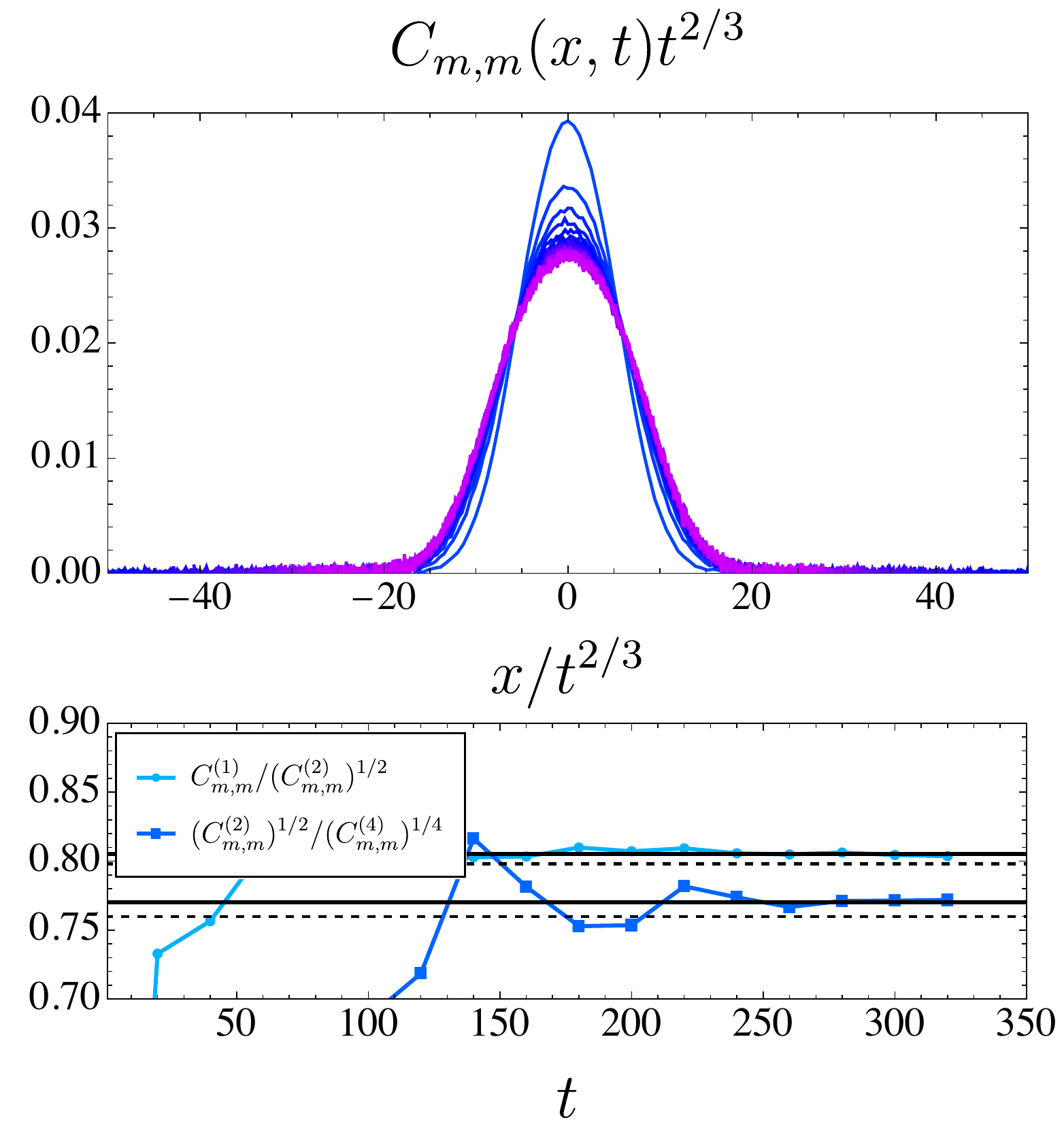}
  \caption{ {\bf KPZ scaling of dynamical correlators of the two-model NLFH.} {\it Top:} Dynamical correlation functions of magnetization $m$ plotted as function of the rescaled space $x/t^{2/3}$ and rescaled by $t^{2/3}$, for different times from $10$ to $400$ with $dt=20$ (from blue to purple). {\it Bottom:} Ratio of spatial moments of the correlation functions $C^{(n)}(t) = \int dx \ C(x,t) \ |x|^n$ as a function of time. The solid ({\it resp.} dashed) lines indicate the universal predictions from $f_{\rm KPZ}(u)$ ({\it resp.} from the Gaussian distribution $f_{\rm Gauss}(u) = e^{-u^2/2\sigma^2}/\sqrt{2 \pi \sigma^2} $.)       }\label{Fig:KPZ correlators}
\end{figure}

 Considering from now on small fluctuations $\delta n$ on top of a reference equilibrium state (i.e. a finite-temperature state with zero net magnetization), and promoting magnetization to be a space and time-dependent function $h \to |m(x,t)|/\chi$, it is simple to check that, see \cite{suppmat}, that \textit{ fluctuations of magnetization are given only in terms of fluctuations of the giant quasi-particles} $ \delta n_{\xi, u } = \lim_{h \to 0} \delta n_{s=\xi/h}(\theta=u/h)$, as 
 \begin{equation} \label{eq:magnefluct}
  {\delta m}{} =    \frac{ -2 m }{\chi} \int d\xi du f_{\xi,u} \delta n_{\xi, u }, 
 \end{equation}
 where $f_{\xi,u}$ a is also a function of  the equilibrium temperature (even in $u$) whose precise definition is irrelevant to us. Therefore, we can simply restrict to the dynamics of their hydrodynamic occupations, reading as 
 \begin{equation}\label{eq:GHDGM}
 \partial_t \delta n _{\xi,u}  + \frac{ |m|}{\chi} \, v^{\rm eff}_{\xi,u} \, \partial_x  \delta n_{\xi,u}=0, 
 \end{equation}
 where we have used that the effective velocity in a thermal state for giant quasiparticles is given by $v^{\rm eff}_{s=\xi/h}(u/h) \to h \ v^{\rm eff}_{\xi,u}$
 with $v^{\rm eff}_{\xi,u}$ a function (odd in $u$) of the two real variables $\xi,u$ as well as a functional of the reference state. Notice the need for the absolute value in the relation between $h$ and $m$, as quasiparticles (i.e. all the conserved quantities of the model except for magnetization) are known to be spin-flip symmetric \cite{Ilievski2016}, as indeed is the case in eq. \eqref{eq:GHDGM}. 
 
 In order to derive the  equation for magnetization we start with the expression of the magnetic current in terms of the velocity of the largest quasiparticle  $v^{\rm eff}_{\infty}$, as 
\cite{piroli2017,dbd2}
\begin{equation}\label{eq:spincurrent}
\partial_t m(x,t)  + \partial_x (v^{\rm eff}_{\infty} m)  =0, 
\end{equation}  
where  $v^{\rm eff}_{\infty} =0$ on any equilibrium state, but it can be finite due to a small fluctuation $\delta n_{\xi,u}$, indeed we have $v^{\rm eff}_{\infty}  = - (2|m|/\chi^2)  \int d\xi du \left( f_{\xi,u} v^{\rm eff}_{\xi,u} \right)   \delta n_{\xi, u}  $, see \cite{suppmat},  which gives 
\begin{equation}\label{eq:GHDGM-2}
\partial_t  \delta m  +  \partial_x \left(  m   \int d\xi du \left( f_{\xi,u} v^{\rm eff}_{\xi,u} \right)  \frac{-2|m|}{\chi^2}  \delta n_{\xi, u} \right) =0,
 \end{equation} 
closing this way the set of hydrodynamic equations necessary to describe magnetic fluctuations.  In order to obtain predictions on the magnetic structure factor $C^{zz}(x,t)$ and the distribution of magnetic fluctuations $P(J_m)$, hydrodynamics must be lifted to fluctuating hydrodynamics. Following the main idea behind non-linear fluctuating hydrodynamics (NLFH)~\cite{PhysRevLett.108.180601,Delfini_2007,spohn_nlfhd,Spohn2014,spohn2012large,Popkov2015,PhysRevLett.120.240601,Jara2015,Lepri2016}, this is achieved by adding noise and dissipation to the currents in \eqref{eq:GHDGM} and \eqref{eq:GHDGM-2}, by respecting fluctuation-dissipation, i.e. stationarity of any Gaussian thermal state of $\delta m$ and all the $\delta n_{\xi,u}$. Clearly solving an infinite set of coupled non-linear stochastic equations poses an immense challenge. In the coming section, we shall reduce the problem to only two modes, an approximation which well captures the emergent KPZ-like physics.

%
%These quasiparticles move with a velocity $v^{\rm eff}_s \sim 1/s  \sim m$, suggesting an intuitive connection with the Burgers equation that was previously pointed out in Refs.~\cite{NMKI19, GVW19}. However, this connection cannot be correct as pointed out above since it would lead to incorrect spin fluctuations, and also because diagonal quadratic terms of the type $j_m = \lambda m^2$ are generically excluded by the structure of normal modes currents in GHD ~\cite{BBH0,1751-8121-50-43-435203,medenjak2019diffusion,doyon2019diffusion}. 

\paragraph{\textbf{Two-mode NLFH } ---}

 The coupled equations \eqref{eq:GHDGM} and \eqref{eq:GHDGM-2}  can be efficiently closed under the approximation 
 \begin{equation}\label{eq:two-mode-approx}
   \int d\xi du  \,  \left( f_{\xi,u} (v^{\rm eff}_{\xi,u})^2  \right) \delta n_{\xi, u }  \propto   \int d\xi du  \,  f_{\xi,u}  \ \delta n_{\xi, u }   ,
\end{equation}  
where the right-hand side is proportional to the magnetic fluctuations \eqref{eq:magnefluct}. Then we can introduce a second field $\delta \phi = ({-2|m|}/{\chi^2} ) \int d\xi du ( f_{\xi,u} v^{\rm eff}_{\xi,u} ) \delta n_{\xi, u} $, and using \eqref{eq:GHDGM}, we obtain two coupled equations for magnetization and field $\phi$,  playing the role of the effective fluid velocity, i.e. $\partial_t \delta m    =    \partial_x   (  \delta m  \delta \phi )$ and the evolution of $\delta \phi$ given by $\partial_t \delta \phi    \propto  \delta |m| \partial_x   (  \delta |m|  )   \propto  \partial_x (\delta m^2)$. 
The approximation of eq. \eqref{eq:two-mode-approx} can be justified by noticing that the square of the velocity $(v^{\rm eff}_{\xi,u})^2$  is even under $u \to - u$, therefore the left hand side of eq. \eqref{eq:two-mode-approx} defines a parity invariant mode, as indeed is the magnetization. The latter is clearly not the only parity invariant mode in general, but the identification is the key to reduce the infinite set of coupled equations  eqs. \eqref{eq:GHDGM}, \eqref{eq:GHDGM-2}  to only a two-mode fluctuating hydrodynamics.  After the inclusion of noise and dissipation, the two-mode NLFH reads 
\begin{align}\label{eq:coupledburger}
\partial_t m  & +  \partial_x \left(   \ m \, \phi  - D_m \partial_x m - \sqrt{2D_m\chi} \xi_m \right) =0  , \notag \\
\partial_t \phi  &  +  \partial_x \left(  \lambda_m   \frac{m^2}{2}+ \lambda_\phi  \frac{\phi^2}{2}  - D_\phi \partial_x \phi - \sqrt{2 D_\phi \chi} \xi_\phi \right)  =0.
\end{align}
The two noise terms in \eqref{eq:coupledburger} are taken to be uncorrelated white noises, and for simplicity, we may choose all diffusion constants and all noise strengths to unity, as well as $\langle m^2 \rangle = \langle \phi^2 \rangle = \chi $ (by appropriate rescaling of the two fields). First, we notice that we can  set $ \lambda_m = 1$ as at this point the Gaussian distribution $ Z^{-1} \exp( {- \frac{1}{2\chi} \int dx \  [m(x,t)^2 + \phi(x,t)^2] })$ is stationary \cite{suppmat}, and as it represent the fixed point of renormalization group (RG) scaling at large $x$ and $t$~\cite{PhysRevLett.69.929} \footnote{Notice that similar equations were first introduced to describe relaxation of drifting polymers ~\cite{PhysRevLett.69.929,PhysRevE.48.1228}, where it was found numerically that the dynamical exponent falls into the KPZ universality class with $z=3/2$.}. As perturbative RG does not seem to be able to fix the coupling $\lambda_\phi$, we will show numerically in the following section, see Fig. \ref{Fig:cumulants}, that at large times  the two eqs. \eqref{eq:coupledburger} converge to two independent Burgers equations with zero velocity and opposite chirality,  
\begin{equation}
\partial_t u_\sigma + \partial_x \left( \sigma \, \frac{u_\sigma^2}{2} - D \partial_x u _\sigma - \xi_\sigma \right)=0, \quad \sigma =\pm,
\end{equation}
  with $u_\sigma \propto m + \sigma \phi$, which we denote as \textit{ two-Burgers decoupling} and which trivially corresponds to the limit $\lambda_\phi \to 1$ of eq. \eqref{eq:coupledburger}.
  
We emphasize that, while we have derived this equation directly from GHD,  eqs.~\eqref{eq:coupledburger} can be entirely fixed by {\em symmetry}: thus \textit{this is the only possible NLFH theory describing spin transport in Heisenberg quantum magnets involving two modes}.  The argument is as follows: first, one identifies $\phi$ as magnetization velocity. Then the current of $\phi$ is fixed by disregarding linear terms (which are zero in \eqref{eq:GHDGM},\eqref{eq:GHDGM-2}) and mixed terms $\sim m \phi$ which are spin-flip odd, while $\phi$ must be spin-flip even.

\paragraph{\textbf{Numerical analysis and two-Burgers decoupling.} ---} We solved numerically the two equations~\eqref{eq:coupledburger} starting from the stationary Gaussian state. We discretize space $x$ with $\Delta x = 1$ on a system of size $L=1000$, implementing a discrete derivative that leaves the Gaussian measure invariant, as explained in~\cite{Sasamoto2009}, and averaging over $\sim 10^5$ realizations, as well as over space (given the translational invariance of the problem). We fix $D_m = D_\phi =1$ and $\chi = 1/2$.  We find that irrespective of the value of $\lambda_\phi$, dynamical correlations follow a KPZ form, i.e. at large times: 
 \begin{align}\label{eq:KPZ correlators}
& C_{m,m}(x,t) =   \langle m  (x,t) m(0,0) \rangle \simeq   \frac{\chi \lambda}{t^{2/3}} f_{\rm KPZ}(\lambda x/t^{2/3}),
\end{align}
and the same for the field $\phi$ (with a different non-universal constant $\lambda$), while $  \langle m  (x,t) \phi(0,0) \rangle  =0$ by symmetry. 
Scaling collapses showing this KPZ behavior are shown in Fig.~\ref{Fig:KPZ correlators}. We also computed ratios of spatial moments of those correlators, and confirmed that they approach the universal prediction from $f_{\rm KPZ}(u)$ at long times (Fig.~\ref{Fig:KPZ correlators}).

We then compute the statistics of the integrated currents  $J = \int_0^t j(x=0,t) dt$ for magnetization and for the chiral mode $\phi$ across the origin $x=0$. In equilibrium, we have no net spin current $\langle J_m \rangle = 0$ (and more generally, all odd cumulants vanish). However, there is a net mean current for the chiral mode $\phi$,  $\langle J_\phi \rangle \sim t$: this ballistic contribution follows from the non-zero average current $\langle j_\phi \rangle =  \langle m^2 \rangle/2 =  \chi/ 2$. Higher even cumulants scale with the dynamical exponent $z=3/2$ as expected: $\langle (J_m)^2 \rangle^c \sim \langle (J_\phi)^2 \rangle^c \sim t^{2/3}$, and $\langle (J_m)^4 \rangle^c \sim \langle (J_\phi)^4 \rangle^c \sim t^{4/3}$ (Fig.~\ref{Fig:cumulants}), where $\langle O^n \rangle^c$ denotes the $n^{\rm th}$ cumulant of $O$. Again, this behaviour is independent of the value of  $\lambda_\phi$.

We then look at the distribution $P_t(J_m)$ of the integrated spin current at time $t$. Introducing the rescaled integrated spin current ${\mathcal J_m} \equiv J_m / t^{1/3}$, we find the universal scaling form
\begin{equation} \label{eqDistribution}
P_t(J_m) \underset{t \to \infty}{\sim} \frac{1}{t^{1/3}}P({\mathcal J_m} \equiv J_m / t^{1/3}), 
\end{equation} 
where the equilibrium probability distribution $P({\mathcal J_m})$ of the rescaled spin current fluctuations is an {\em even} function (as it should). Our theory thus reconciles KPZ scaling of the spin dynamical correlation function~\eqref{eq:KPZ correlators} with a symmetric distribution of spin fluctuations~\eqref{eqDistribution}.
Moreover, we find that also the latter is independent of the value of $\lambda_\phi$, see Fig. \ref{Fig:cumulants}, as shown by its kurtosis that for different values of $\lambda_\phi$ converges to the one given by the sum of two independent Burgers equations with opposite chirality (corresponding to the case $\lambda_\phi = 1$). In this case, the distribution of the sum of the two modes is the convolution of two Baik-Rains, and therefore with kurtosis ${\rm Ku}  ={\rm Ku}_{\rm BR} /2  $ with ${\rm Ku}_{\rm BR} \sim 0.28$. Therefore we conclude that the two-mode theory of eq. \eqref{eq:coupledburger} is equivalent to two Burgers fields with opposite chirality, characterized by KPZ two-point functions and Baik-Rains fluctuations with opposite skewness.

  \begin{figure}[t!]
  \includegraphics[width=0.48 \textwidth]{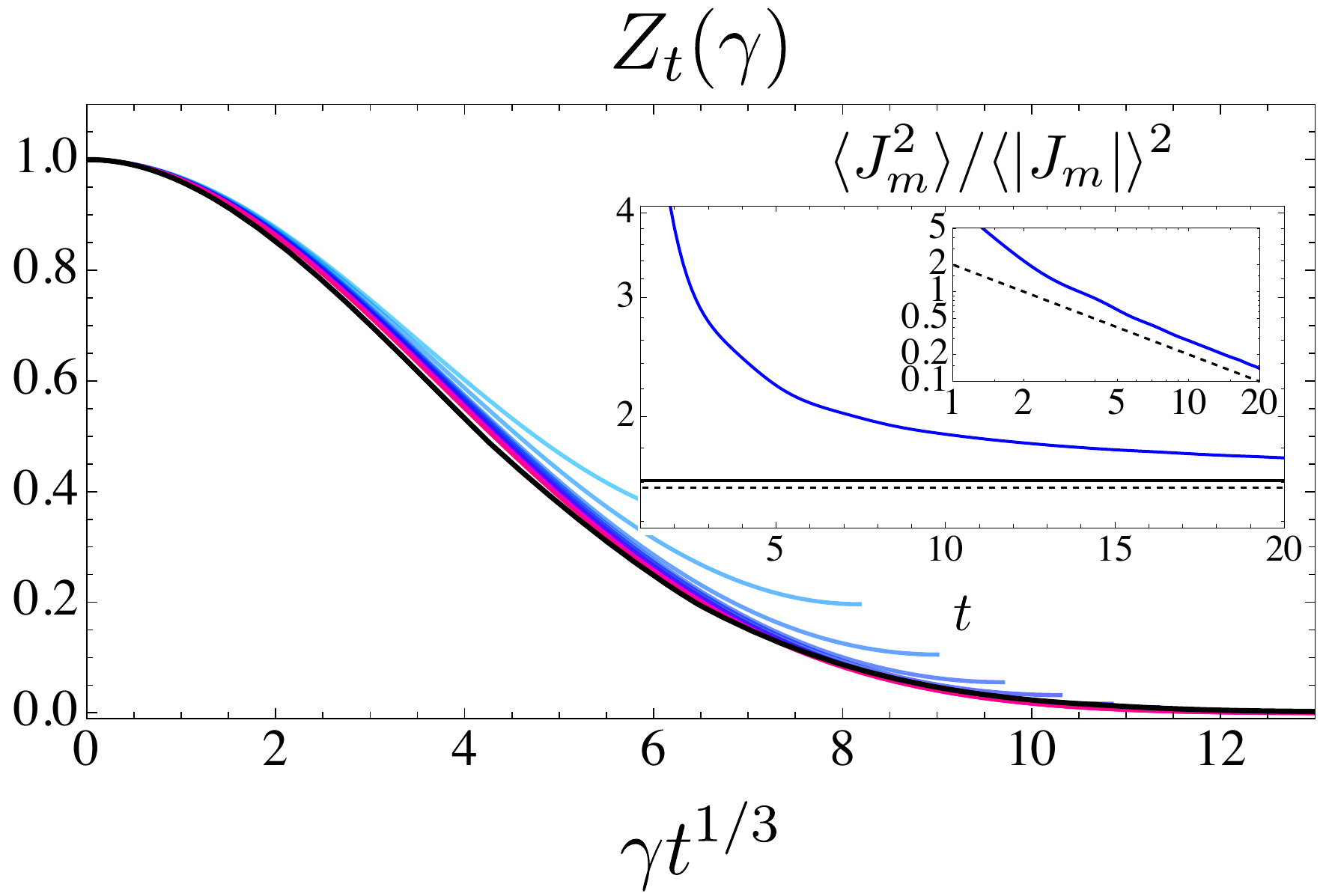}
  \caption{ {\bf Moment generating function of the integrated spin current.}  Plot of $Z_t(\gamma)$, eq. \eqref{eqFCSnumerics}, using MPS (coloured lines) evaluated at different times $t =0-20$ (from light blue to purple) plotted as a function of $\gamma t^{1/3}$, compared with the prediction of Eqs~\eqref{eq:coupledburger} (black thick line) at late times. The scale of the $x$ axis is set by fixing the same second derivative at $\gamma=0$ for MPS and theory.  Inset: Plot of the cumulant $\langle J^2_m \rangle/\langle | J_m | \rangle^2$ (where $ | \cdot |$ denotes absolute value) obtained from MPS versus the prediction of two-mode NLFH and the prediction of Gaussian distribution ($\pi/2$) (dashed black  line). Inset of Inset: log-log plot of the difference as function of time versus the line $t^{-1}$.      }\label{Fig:FCS}
\end{figure}

\paragraph{\textbf{Matrix-product states computation} ---} In order to substantiate our predictions, we compute numerically the statistics of spin current fluctuations using matrix-product states (MPS) techniques~\cite{PhysRevLett.91.147902,whitetdmrg,schollwoeck}, see~\cite{PhysRevLett.107.206801} for a similar calculation in the context of interacting quantum dots. To keep track of spin transport, we rewrite the Hamiltonian density as $ \frac{1}{2} ( t_{xy} S^+_j S^-_{j+1} + {t}^*_{xy}  S^-_j S^+_{j+1}) +  S^z_j S^z_{j+1}$ with $t_{xy} = 1$, and introduce a counting field $\gamma$ via the substitution $t_{xy} \to {\rm e}^{i \gamma/2}$ on the central bond of the system. Denoting the resulting modified Hamiltonian $H_\gamma$, the following overlap is precisely the generating function $Z_t(\gamma) $  of the integrated spin current $J_m = \int_0^t dt j_m(x=0,t)$ (using standard prescriptions of time-ordering of the measurements~\cite{doi:10.1063/1.531672}):
\begin{equation} \label{eqFCSnumerics}
Z_t(\gamma) = \frac{1}{2^L}{\rm Tr} \left( {\rm e}^{i H_{-\gamma} t} {\rm e}^{-i H_{\gamma} t} \right) =  \langle e^{i \gamma J_m} \rangle.
\end{equation} 
Computing the generating function $Z_t(\gamma)$ up to time $t=20$ with maximal bond dimension $800$ on a chain of size $L=100$, we find a universal collapse versus the rescaled variable $\gamma t^{1/3}$, in agreement with eq.~\eqref{eqDistribution} (Fig.~\ref{Fig:FCS}). The MPS results are compatible with a distribution very close to a Gaussian, in agreement with our predictions, with a slow $t^{-1}$ approach to the non-linear fluctuating hydrodynamic behavior. 

\paragraph{\textbf{Discussion} ---}
In this letter, we have developed a scenario for superdiffusion in the Heisenberg spin chain that reconciles the observed KPZ scaling functions with the presence of inversion symmetry. Our central equations~\eqref{eq:coupledburger} are derived from generalized hydrodynamics, but they are also strongly constrained by symmetry: in fact, Eqs.~\eqref{eq:coupledburger} are the most general two-mode hydrodynamic equations consistent with the symmetries of the Heisenberg model. We have shown that these two equations decouple at larger times to two Burgers equations with opposite chirality, with magnetization given by the sum of the two fields. 
While the two-mode theory is an approximation, it remains to understand how much it captures of the full system of infinite coupled equations  eqs. \eqref{eq:GHDGM}, \eqref{eq:GHDGM-2}. One possible scenario is that the two-Burgers decoupling we find in the two-mode theory also exists for a much larger number of coupled equations, leading then to magnetization fluctuations resulting from the sum of an infinite set of Burgers fields, giving therefore Gaussian statistics at late times (which would be more compatible with finite-time data from numerical simulations \cite{PhysRevLett.128.090604,2303.16691,2306.09333}). As the analytical and numerical solution of such an infinite set of stochastic equations is very challenging, we hope to come back to this question in the future.

Nevertheless, we emphasize that based on existing numerical results, our two-mode hydrodynamics captures all the numerically observed features. In particular, our equations reproduce KPZ scaling for the dynamical spin structure factor, while predicting a universal symmetric distribution for current fluctuations that is consistent with direct numerical studies. The distribution we predict is close to a Gaussian one with a small kurtosis, as Refs. \cite{PhysRevLett.128.090604,2303.16691,2306.09333} also found, and our result of Fig. \ref{Fig:cumulants} moreover shows that the crossover to two independent Burgers can take very long time, approaching from Gaussian fluctuations. Another important application of our hydrodynamics would be to study the effects of integrability breaking perturbations that preserve the $SU(2)$ symmetry: numerical evidence and perturbative calculations suggest that superdiffusion is more stable than dimensional analysis would predict~\cite{PhysRevLett.124.210605, PhysRevLett.127.057201, glorioso2021hydrodynamics, PhysRevB.105.L100403, roy2022robustness, mcroberts2022long}, and the rigidity of our hydrodynamic framework might provide some insight into this stability. Finally, the counting field expression~\eqref{eqFCSnumerics} might be a useful starting point for analytic techniques using integrability.

\paragraph{\textbf { Acknowledgements -- }} We thank Immanuel Bloch, Vir Bulchandani, Enej Ilievski, Vedika Khemani, Ziga Krajnik, Ewan McCulloch,  Alan Morningstar, Tomaz Prosen, Andrea De Luca, Pierre Le Doussal for helpful discussions. We also thank Herbert Spohn, for discussions and for kindly pointing out references~\cite{PhysRevLett.69.929} and~\cite{Sasamoto2009} to us, and Andrea Sportiello for helpful suggestions at the Galileo Galilei Institute during the scientific program “Randomness, Integrability, and Universality”, and Kazumasa Takeuchi for the ongoing collaboration.  Some of this work was performed
at Aspen Center for Physics, which is supported by National Science Foundation grant
PHY-1607611. This work was supported by the ERC Starting Grant 101042293 (HEPIQ) (J.D.N.), the National Science Foundation under NSF Grant No. DMR-1653271 (S.G.),  the Air Force Office of Scientific Research under Grant No. FA9550-21-1-0123 (R.V.), and the Alfred P. Sloan Foundation through a Sloan Research Fellowship (R.V.).

\bibliography{SSD,refs,biblio}

%\bigskip
%
%\onecolumngrid
%\newpage
%
%\includepdf[pages=1]{SM.pdf}
%\newpage
%\includepdf[pages=2]{SM.pdf}
%\newpage
%\includepdf[pages=3]{SM.pdf}
 
\newpage

 \onecolumngrid

\begin{center}
\textbf{\large Supplementary Material} \\ 
\end{center}
\label{appmain}

\section{Invariance of the Gaussian measure under coupled KPZ equations}

We here show that  Gaussian measure 
\begin{equation}
P =Z^{-1} \exp\left( {- \frac{1}{2\chi} \int dx \  \left[(\partial_x h_m)^2 + (\partial_x h_\phi)^2 \right] } \right),
\end{equation}
is stationary under the coupled KPZ equations
\begin{align}\label{eq:coupledKPZ}
\partial_t h_m + \lambda_m \ \partial_x h_m  \partial_x h_\phi &=  D_m \partial^2_x h_m + \xi_m, \\
\partial_t h_\phi  +\frac{\lambda_m}{2}    \left(   \partial_x h_m     \right)^2 &= D_\phi \partial^2_x h_\phi + \xi_\phi,
\end{align}
 where we have introduced the height fields $m(x,t) = \partial_x h_m$ and $\phi(x,t) = \partial_x h_\phi$. 
The stationary measure of those equations satisfies
\begin{equation}
    \sum_{i=m,\phi} \int dx \frac{\delta}{ \delta h_i(x)} \left( -D_i \partial^2_x h_i  + G^{kl}_{i} (\partial_x h_k) (\partial_x h_l)  +  \frac{1}{2} \frac{\delta }{ \delta h_i(x)} \right) P =0 ,
\end{equation}
where $G^{m\phi}_{m}=G^{\phi m}_{m} = \lambda_m/2$, $G^{m m}_{\phi} = \lambda_m/2$ characterize the non-linear terms in~\eqref{eq:coupledKPZ}. This equation relates the strength of the noise and dissipation to the susceptibility $\chi $, which we solve by setting $D_m=D_\phi=1$ and taking the noise to have unit strength so that $\chi=1/2$. Meanwhile, the nonlinear terms lead to the condition
\begin{equation}
    \int dx  \left( \frac{\delta}{\delta h_m(x)} [\lambda_m (\partial_x h_m) (\partial_x h_\phi) ] P +  \frac{\delta}{ \delta h_\phi(x)} [\frac{\lambda_m}{2} (\partial_x h_m)^2   ] \right)P=0,
\end{equation}
which is satisfied by the Gaussian measure since
\begin{equation}
  - \lambda_m \int dx   [ (\partial_x h_m) (\partial_x h_\phi) ] \partial_x^2 h_m - \frac{\lambda_m}{2} \int dx   [ (\partial_x h_m)^2   ] \partial_x^2 h_\phi = - \frac{\lambda_m}{2} \int dx \frac{\partial }{\partial x} \left[ (\partial_x h_m)^2  \partial_x h_\phi \right]=0,
\end{equation}
up to boundary terms.

\section{Additional details on the giant quasiparticle GHD derivation}

{ In this appendix, we derive our two NLFH equations directly from GHD. We will use extensively the Thermodynamic Bethe Ansatz (TBA) solution of the Heisenberg quantum spin chain, and refer the reader to Ref.~\cite{Takahashi} for details. }
Let us start by deriving the first equation for the magnetization $m$ in the Heisenberg chain at Euler scale
\begin{equation}
    \partial_t m(x,t)  + \lambda_m   \partial_x ( m(x,t)  \phi (x,t)  ) = 0.
\end{equation}
From GHD, we have the following equation at Euler scale for the magnetization 
\begin{equation}
    \partial_t m(x,t) + \partial_x (m(x,t) v^{\rm eff}_\infty(x,t)) =0.
\end{equation}
The quantity $v^{\rm eff}_\infty = \lim_{s \to \infty} v_s^{\rm eff}$ depends non-trivially on the global equilibrium state, specified by chemical potentials $\{\beta^i\}_i$ (in general there is an infinite number of them, as the model is integrable) via the thermodynamic Bethe ansatz dressing, and it is zero for any finite temperature equilibrium state. We consider therefore a small, space dependent, perturbation around a (infinite) temperature state $\beta^i   = \beta \delta_{i,1}$, namely, using repeated indices convention, 
\begin{equation}
    v^{\rm eff}_\infty(\beta + \delta \beta^i) =  \delta \beta ^i \frac{\partial v^{\rm eff}_\infty}{\partial \beta^i}   =  
     \frac{\partial v^{\rm eff}_\infty}{\partial n_j}  \delta n_j  , 
\end{equation}
where we used the collective index $j$ to denote strings $s$ and momenta $u$. We have introduced the filling functions $n_s(u)$ characterizing the equilibrium state with generic chemical potentials, so that we have $n_j(\beta_0^i + \delta \beta^i) = n_j + \delta n_j$, with $\delta n_j = (\partial n_j /\partial \beta^i) \delta \beta^i$. Moreover, the total densities of states $k'_s(u) = 2 \pi \rho_s^{\rm tot}(u)$ and the velocities are $v^{\rm eff}_s(u)$, for each type $s$ of quasiparticle. This also allows us to define the dressing operation $f^{\rm dr} = (1- T n)^{-1} f$ as a linear operation on any (integrable) function, in terms of the scattering shift $T_{s,s'}(u,u')$ of the model. 
From generic dressing relations, see for example~\cite{Doyon_notes},  we have, with $\chi$ the spin susceptibility, 
\begin{equation}
 \delta f^{\rm dr}_j =    T^{\rm dr}_{j,j'} f^{\rm dr}_{j'} \delta n_j , \quad  \quad   \delta  v^{\rm eff}_\infty =- 2 \frac{v^{\rm eff}_j \rho_j^{\rm tot} \mu^{\rm dr}_j}{ \chi} \delta n_j.
\end{equation}
Here, the dressed magnetic moment is obtained as the low field limit of the dressed magnetization $\mu^{\rm dr}_j = \lim_{h \to 0^+} m^{\rm dr}_j/h$, where $h$ is a global magnetic field $\chi h= |m|$, but also from the large $s$ limit of the dressed scattering kernel
\begin{equation}
- 2\mu^{\rm dr}_s/\chi =  \lim_{s' \to \infty} T^{\rm dr}_{s',s}/\rho^{\rm tot}_{s'} ,
\end{equation}
Notice that since the velocities $v^{\rm eff}_j$ on an equilibrium state are all odd functions of rapidity $\theta$, the only modifications that give non-zero contributions are the antisymmetric in momenta $u$. Restoring summation over $s$ and integration over $u$ and considering fluctuations with identical susceptibility, i.e. by rescaling $\delta n_i \to \sqrt{\chi_i} \delta n_i$ with $\chi_i = \rho_i (1-n_i)$, we obtain:
\begin{equation}
   \delta v^{\rm eff}_\infty =  \frac{-2}{ \chi}  \sum_s \ \int_{-\infty}^\infty d\theta \ \Big[ \rho^{\rm tot}_s(\theta) v^{\rm eff}_s(\theta) \mu^{\rm dr}_s  \sqrt{\chi_s(\theta)} \Big]_{\beta} \ \delta n_{s}(\theta) ,
\end{equation}
where the block $[\cdot]_\beta$ is evaluated at finite (or infinite) temperature and is a functional of the global equilibrium state.
We now can consider the following semi-classical giant quasiparticle rescaling 
\begin{equation}
    s = \xi /h ; \theta = u / h ,
\end{equation}
In this way we have the following scaling, see \cite{NGIV20}
\begin{equation}
    \mu_s \to h^{-2} \mu_{\xi,u}, \quad\sqrt{\chi_s}  \to h^2  \sqrt{\chi_{\xi,u}}, \quad \rho^{\rm tot}_s(\theta) \to h^2 \rho^{\rm tot}_{\xi,u}, \quad  v^{\rm eff}_s(\theta) \to h v^{\rm eff}_{\xi,u}.
\end{equation}
 Therefore, in the giant quasiparticles scaling, using $\sum_s \int d\theta \to h^{-1} \int d\xi \int du$, 
\begin{equation}
  \delta v^{\rm eff}_\infty =   \frac{-2}{ \chi}  \sum_s \ \int_{-\infty}^\infty d\theta \ \Big[ \rho^{\rm tot}_s(\theta) v^{\rm eff}_s(\theta) \mu^{\rm dr}_s  \sqrt{\chi_s(\theta)} \Big]_{\beta} \ \delta n_{s}(\theta)  \to   \frac{-2}{ \chi} \int_0^\infty d\xi \int_{-\infty}^{+\infty} du  \ v^{\rm eff}_{\xi,u} f_{\xi, u} v^{\rm eff}_{\xi,u} \ h \delta n_\xi(u),
\end{equation}
where the function
\begin{equation}
 f_{\xi, u}  =  \mu_{\xi,u} \rho^{\rm tot}_{\xi,u} \sqrt{\chi_{\xi,u}}. 
\end{equation}
only depends on the reference equilibrium state around which the modes are fluctuating. Finally, using $h = |m|/\chi$ we have 
\begin{equation}
  \delta v^{\rm eff}_\infty =    \frac{-2 |m| }{ \chi^2} \int_0^\infty d\xi \int_{-\infty}^{+\infty} du  \ v^{\rm eff}_{\xi,u} f_{\xi, u} v^{\rm eff}_{\xi,u} \  \delta n_\xi(u),
\end{equation}
Next we consider magnetic fluctuations in terms of quasiparticles. It is well-known that quasiparticles are spin flip invartiant and therefore they can only fix the absolute value of the magnetization. This is given by 
\begin{equation}
|m| =\frac{1}{2}  \lim_{s \to \infty} \int   d\theta \ \rho^{\rm tot}_s(\theta),
\end{equation}
and therefore the fluctuations of the absolute value of the magnetizations are given by
\begin{equation}
  \delta   |m| =  \frac{1}{2} \lim_{s \to \infty} \int d\theta \delta  \rho_s^{\rm tot}(\theta) =  \frac{1}{2} \lim_{s \to \infty} \int d\theta \rho^{\rm tot}_s(\theta)  \sum_{s'} \int d\theta'  \frac{-2\mu_{s'}}{ \chi} \rho^{\rm tot}_{s'}(\theta') \sqrt{\chi_{s'}(\theta')}  \delta n_{s'}(\theta') \to   - \frac{2|m|}{\chi}  \int du d\xi \ f_{\xi,u} \delta {n}_{\xi,u},
\end{equation}
and given that $\delta |m| = {\rm sgn}(m) \delta m + m \delta {\rm sgn}(m)$ and that $m \delta {\rm sgn}(m) =0$, we obtain 
\begin{equation}
  \delta   m =    - \frac{2m}{\chi}  \int du d\xi \ f_{\xi,u} \delta {n}_{\xi,u}. 
\end{equation}

\end{document}